\documentclass[conference, 10pt]{IEEEtran}

\usepackage{cite}
\usepackage{url}
\usepackage{graphicx}
\usepackage{color}
\usepackage{placeins}
\usepackage{float}
\usepackage{tabularx,colortbl}
\usepackage{ifthen}
\usepackage[font={small}]{caption}
\usepackage[font={small}]{subcaption}
\usepackage{overpic}

\makeatletter

\newcounter{author}
\renewcommand{\author}[2][]{
   \stepcounter{author}
   \@namedef{author@\theauthor}{#2}
   \@namedef{authorlabel@\theauthor}{#1}
}

\newcounter{address}
\newcommand{\address}[2][]{
   \stepcounter{address}
   \@namedef{address@\theaddress}{#2}
   \@namedef{addresslabel@\theaddress}{#1}
}

\newcommand{\alsep}{and}

\def\newmaketitle{\par%
  \begingroup%
  \normalfont%
  \def\thefootnote{}%  the \thanks{} mark type is empty
  \def\footnotemark{}% and kill space from \thanks within author
  \let\@makefnmark\relax% V1.7, must *really* kill footnotemark to remove all \textsuperscript spacing as well.
  \footnotesize%       equal spacing between thanks lines
  \footnotesep 0.7\baselineskip%see global setting of \footnotesep for more info
  \normalsize%
  \twocolumn[\thenewmaketitle\@IEEEaftertitletext]%
  \if@IEEEusingpubid
     \enlargethispage{-\@IEEEpubidpullup}%
  \fi
  \endgroup
  \setcounter{footnote}{0}\let\maketitle\relax\let\@maketitle\relax
  \gdef\@thanks{}%
  \let\thanks\relax}

\def\thenewmaketitle{
  \newpage
  \begin{center}%
    \vskip0.2em{\Huge\@IEEEcompsoconly{\sffamily}\@IEEEcompsocconfonly{\normalfont\normalsize\vskip 2\@IEEEnormalsizeunitybaselineskip
   \bfseries\large}\@title\par}\vskip1.0em\par%
    \vspace{1ex}
    \newcounter{c@author}
    \newcounter{c@tmp}
    \ifthenelse{\value{author}=2}{%
      \newcommand{\liand}{ and }}{%
      \newcommand{\liand}{, and }}
    \ifthenelse{\value{address}<2}{%
      \@nameuse{author@1}%
      \stepcounter{c@author}%
      \whiledo{\value{c@author}<\value{author}}{%
        \setcounter{c@tmp}{\value{author}}%
        \addtocounter{c@tmp}{-\value{c@author}}%
        \ifthenelse{\value{c@tmp}=1}{%
          \renewcommand{\alsep}{\liand}}{\renewcommand{\alsep}{, }}%
        \stepcounter{c@author}\alsep \@nameuse{author@\thec@author}}\\%
    }
    {%Add address references after the author's name
      \@nameuse{author@1}${}^{(\ref{\@nameuse{authorlabel@1}})}$%
      \stepcounter{c@author}%
      \whiledo{\value{c@author}<\value{author}}{%
      \setcounter{c@tmp}{\value{author}}%
      \addtocounter{c@tmp}{-\value{c@author}}%
      \ifthenelse{\value{c@tmp}=1}{%
        \renewcommand{\alsep}{\liand}}{\renewcommand{\alsep}{, }}%
      \stepcounter{c@author}\alsep \@nameuse{author@\thec@author}%
        ${}^{(\ref{\@nameuse{authorlabel@\thec@author}})}$%
      }
    }
    \vspace{0.2ex}

    \ifthenelse{\value{address}>0}{%
      \ifthenelse{\value{address}=1}{
        {\@nameuse{address@1}}
      }
      {%Output the addresses as an enumerated list
        \newcounter{c@address}

        \begin{center}
        \whiledo{\value{c@address}<\value{address}}
        {
          \refstepcounter{c@address}
            ${}^{(\thec@address)}$\,%
              \label{\@nameuse{addresslabel@\thec@address}}%
              \@nameuse{address@\thec@address}\\ %
        }
        \end{center}
      } % end of the address creation ifthenelse block
    }
    {
      \relax
    }
  \end{center}
}

\makeatother

\title{Computationally-efficient synthesis of inversely-designed  3D-printable all-dielectric devices }

\author[1]{Maria-Thaleia Passia}
\author[1]{Steven A. Cummer}

\address[1]{Department of Electrical and Computer Engineering, Duke University, NC, USA}

\let\svthefootnote\thefootnote

\newcommand\blankfootnote[1]{%
  \let\thefootnote\relax\footnotetext{#1}%
  \let\thefootnote\svthefootnote%
}

\begin{document}

\newmaketitle

\begin{abstract}
  We present a systematic, computationally efficient approach for synthesizing 3D-printable all-dielectric devices.  Inverse-design optimization methods lead to  devices  of a continuous dielectric constant profile with complex and conformal shapes. However, stereolithography 3D printers have a limited range of materials; usually, only resin and air are available.  As the size and complexity of the devices increase, performing simulations of the entire detailed manufacturable device becomes computationally challenging or even prohibitive. We introduce the \texttt{LOCABINACONN} methodology for transforming an optimized device of a continuous material profile to a manufacturable one while preserving performance as close as possible to the continuous case. The \texttt{LOCABINACONN} is a local and computationally efficient methodology where we identify suitable air/resin configurations that will substitute non-manufacturable material components without simulating the entire manufacturable device. This work paves the way for synthesizing optimized larger-scale 3D-printable devices in a computationally tractable manner.
\end{abstract}

\section{Introduction}
\IEEEPARstart{A}{dvances} in additive manufacturing have enabled the realization of easily deployable devices with rapid prototyping. Among various additive manufacturing techniques, resin-based, as is stereolithography (SLA) 3D-printing, provide the potential to fabricate structures of even higher precision. Various all-dielectric devices, fabricated by SLA 3D-printing, have been reported in recent literature, such as a multibeam lens antenna for CubeSat applications~\cite{Trzebiatowski2022}, a flat Luneburg lens antenna using ceramic SLA 3D-printing~\cite{Lou2021}, a THz metalens~\cite{Jang2023}.

\blankfootnote{Copyright (C) 2025 IEEE. Personal use of this material is permitted.  Permission from IEEE must be obtained for all other uses, in any current or future media, including reprinting/republishing this material for advertising or promotional purposes, creating new collective works, for resale or redistribution to servers or lists, or reuse of any copyrighted component of this work in other works.}

Inverse-design optimization techniques lead to all-dielectric devices of very high efficiency. Topology optimization techniques, such as objective-first optimization~\cite{Lu2012} and adjoint optimization approaches~\cite{Sell2017}, have been used to design a variety of nanophotonic and millimeter-wave devices~\cite{Piggott2014,Callewaert2016}. Commonly, inverse design approaches yield devices with a continuous range of dielectric constant values. However, the available materials for SLA 3D printing are usually limited to a single resin. Fabrication constraints are, in many cases, introduced in the inverse design optimization~\cite{Vercruysse2019,Callewaert2016} to promote manufacturable devices. A device with a continuous dielectric constant profile must be transformed into a manufacturable one while preserving performance as close as possible to the continuous case. 

A common approach for transforming a non-manufacturable all-dielectric device to a manufacturable one is to substitute the materials that cannot be manufactured by suitable air/resin configurations of a predetermined shape~\cite{Xie2018,Ahmed2023,Xie2023}. In the case of inversely designed devices, materials form detailed complex and conformal components that vary across the device. Hence, finding a single air/resin structure of a predetermined shape that is suitable for all components is not feasible.  Assigning a resin percentage to each material instead presents a more feasible alternative. Suitable air/resin configurations that approach the response of the non-manufacturable material layers must be identified. The  selection of  suitable air-resin structures can be performed on a device level, requiring simulations of the entire manufacturable device~\cite{Passia2024}.
However, as devices increase in size and complexity, performing simulations of the entire detailed manufacturable device becomes computationally-challenging or even prohibitive.

In this paper, we introduce the \texttt{LOCABINACONN} methodology to synthesize inversely-designed manufacturable devices by identifying  suitable air/resin configurations on a local component level. We produce several manufacturable air/resin configurations of a prescribed resin percentage for each smaller material component and determine the propagation characteristics of each composite configuration. We select the manufacturable configuration which better approaches the propagation characteristics of the non-manufacturable material component. To demonstrate our approach, we synthesize a diffraction metagrating that preserves performance very close to the continuous case. By simulating only smaller material components, our methodology paves the way for synthesizing larger-scale optimized 3D-printable devices in a computationally-tractable manner, suitable for microwave and mmWave applications.

\section{Overview of the \texttt{LOCABINACONN} methodology}

\begin{figure}
  \centering
\includegraphics[width = 1\columnwidth,trim={0cm 0cm 0cm 0cm},clip]{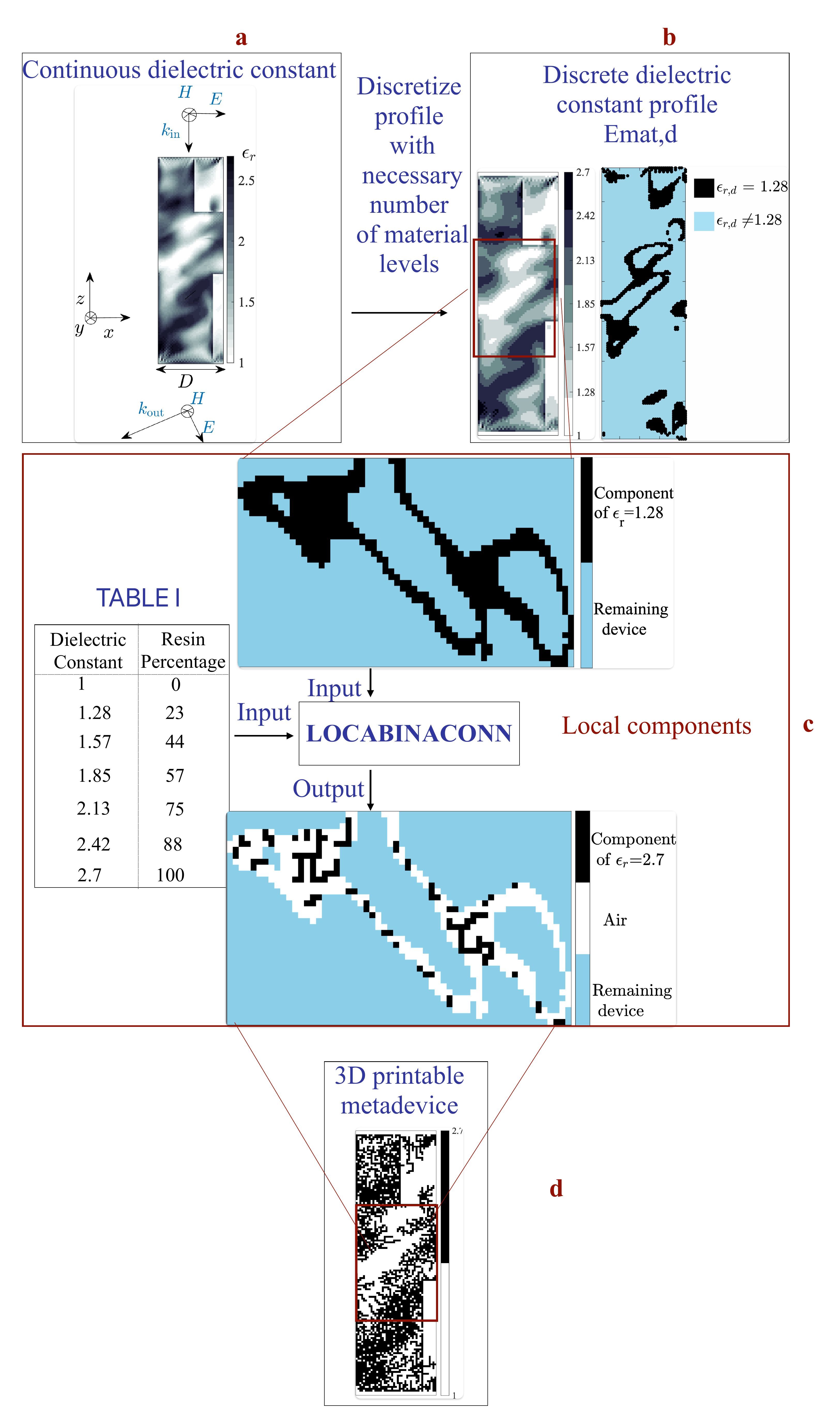}\\
\caption{Overview of the \texttt{LOCABINACONN} methodology}
\label{fig:overview}
\end{figure}

We present an overview of the \texttt{LOCABINACONN} methodology in Fig.\ref{fig:overview}. The  \texttt{LOCABINACONN} methodology takes an inversely designed continuous device as an input Fig.\ref{fig:overview}a and transforms it to a manufacturable one Fig.\ref{fig:overview}d, by replacing non-manufacturable components with manufacturable ones on a local  level.  
In order to demonstrate our methodology, we consider, as a proof of concept, a diffraction metagrating, which refracts a normally incident plane wave from the top, with the magnetic field oriented along the $y$-axis, following the axis convention of Fig.~\ref{fig:overview},  to a wide angle of $-77^o$. The metagrating is assumed to be infinite in extent along the $y$ axis. As the device is periodic along the $x$ axis, having a period $D$, we examine the device's unit cell.

The inversely-designed device with a continuous dielectric constant profile is shown at the upper left corner of Fig.\ref{fig:overview}, and is termed as Fig.\ref{fig:overview}a. 
This device has a continuous dielectric constant profile in the range $\epsilon_r=\{1,2.7\}$, where $\epsilon_r=1$  corresponds to air and $\epsilon_r=2.7$ corresponds to an SLA 3D-printer resin, such as Formlabs Clear V4 resin, characterized in~\cite{Palazzi2019}.  We first transform the device of a continuous profile to a discrete one, by keeping as many material levels as needed to maintain the efficiency of the continuous device. The discrete device is termed device Fig.\ref{fig:overview}b, and is also a non manufacturable device. 

The \texttt{LOCABINACONN} methodology has two inputs: 1.  smaller non-manufacturable components of device  Fig.\ref{fig:overview}b, as is the pattern with $\epsilon_r$ = 1.28 in Fig.\ref{fig:overview}c, and 2. a table that assigns a resin percentage to each dielectric constant value. The output of \texttt{LOCABINACONN} is a manufacturable component, composed of only resin and air. The entire manufacturable device unit cell is synthesized by assemblying all smaller manufacturable components.

\section{The \texttt{LOCABINACONN} methodology}

\subsection{Inversely designed device of a continuous dielectric constant profile}
\begin{figure}
  \centering
    \subfloat[]{\label{fig:cont_geom_eff}%
    \includegraphics[width = 0.8\columnwidth]{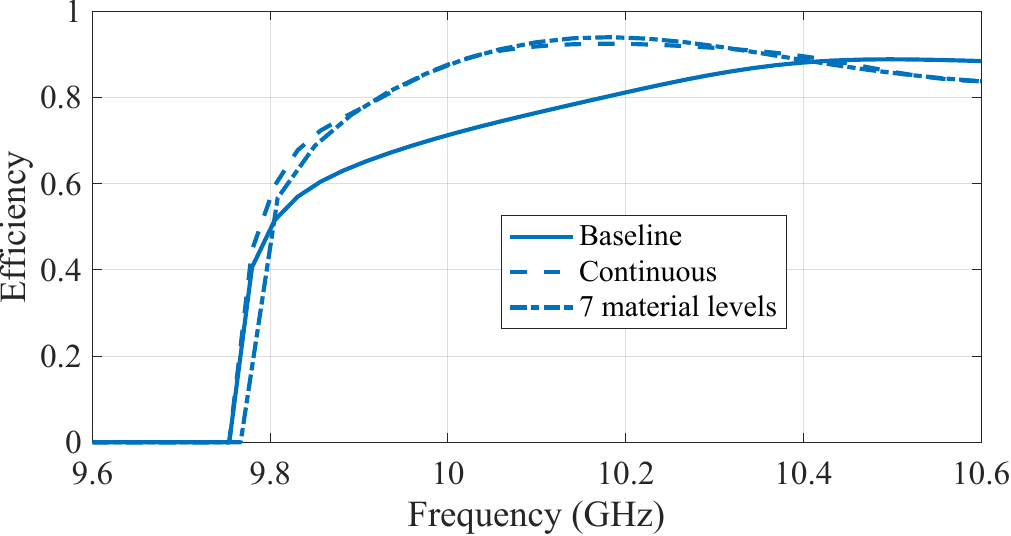}
    }
    \subfloat[]{\label{fig:base_geom}%
    \includegraphics[width = 0.19\columnwidth]{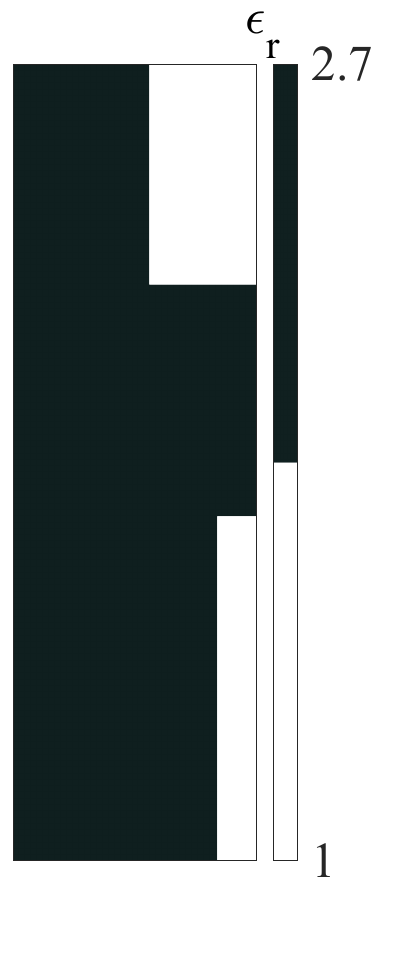}
      }   
    \caption{(a) The diffraction efficiency of the -1-diffraction order for the baseline device, the optimized device of a continuous material profile, and  the optimized device of a discrete but non-manufacturable profile.
    (b) Baseline metagrating.}
\end{figure}

The metagrating of a continuous dielectric constant profile is obtained by  objective-first inverse design optimization~\cite{Lu2012} (Fig.\ref{fig:overview}a). The objective-first optimization was implemented on a 141$\times$43 finite difference uniform square grid, with each cell being of size $\lambda/42\times \lambda/42$, where $\lambda$ is the wavelength at 10 GHz.  We set the propagation properties of a normally incident plane wave as input, and those of a wave diffracted into a $-77^o$ angle as output. By starting with a baseline geometry of a diffraction grating of simpler geometric shape~\cite{Dong2020} and the same materials and dimensions, we obtained the optimized device of Fig.~\ref{fig:overview}, with each cell of the grid having a value between $\epsilon_r=1$ and $\epsilon_r=2.7$. The $-1$ diffraction order efficiency of both the metagrating with a continuous dielectric constant profile and the baseline  geometry is shown in Fig.~\ref{fig:cont_geom_eff}.  We observe that the optimized continuous device has a greater diffraction efficiency than the baseline configuration within the [9.8,10.4] GHz range.

%\vspace{-0.5em}

\subsection{Transforming a device of a continuous dielectric constant profile to a discrete one}
To determine the number of discrete material levels, we perform FEM simulations of the device with a different number of discrete levels retained in each simulation. All continuous dielectric constant values are substituted by the closest discrete value.  By adding more intermediate levels, the performance approaches to that of the continuous case.   We select the device with a discrete dielectric profile (Fig.\ref{fig:overview}b), which achieves performance as close as possible to that of the continuous case and for which adding more material levels offers a negligible improvement in the performance. Seven material levels are sufficient in the considered example, as verified by comparing the $-1$ diffraction order efficiency of the device with seven material levels to that of the continuous profile in Fig.~\ref{fig:cont_geom_eff}.

\begin{figure} 
  \centering
\begin{overpic}[width=0.9\linewidth,trim = 0cm 0cm 0cm 0cm,clip]{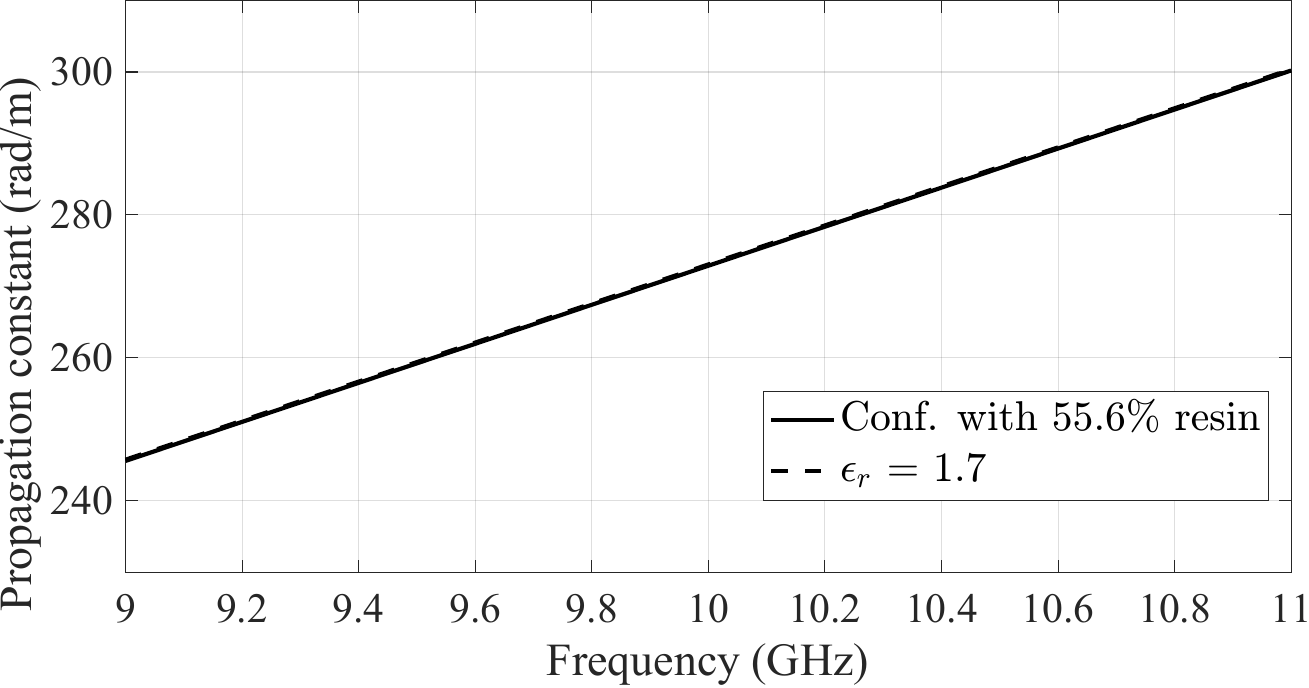}
  \put(14,31){
      \includegraphics[width=0.3\linewidth,trim = 0cm 1cm 0cm 0cm,clip]{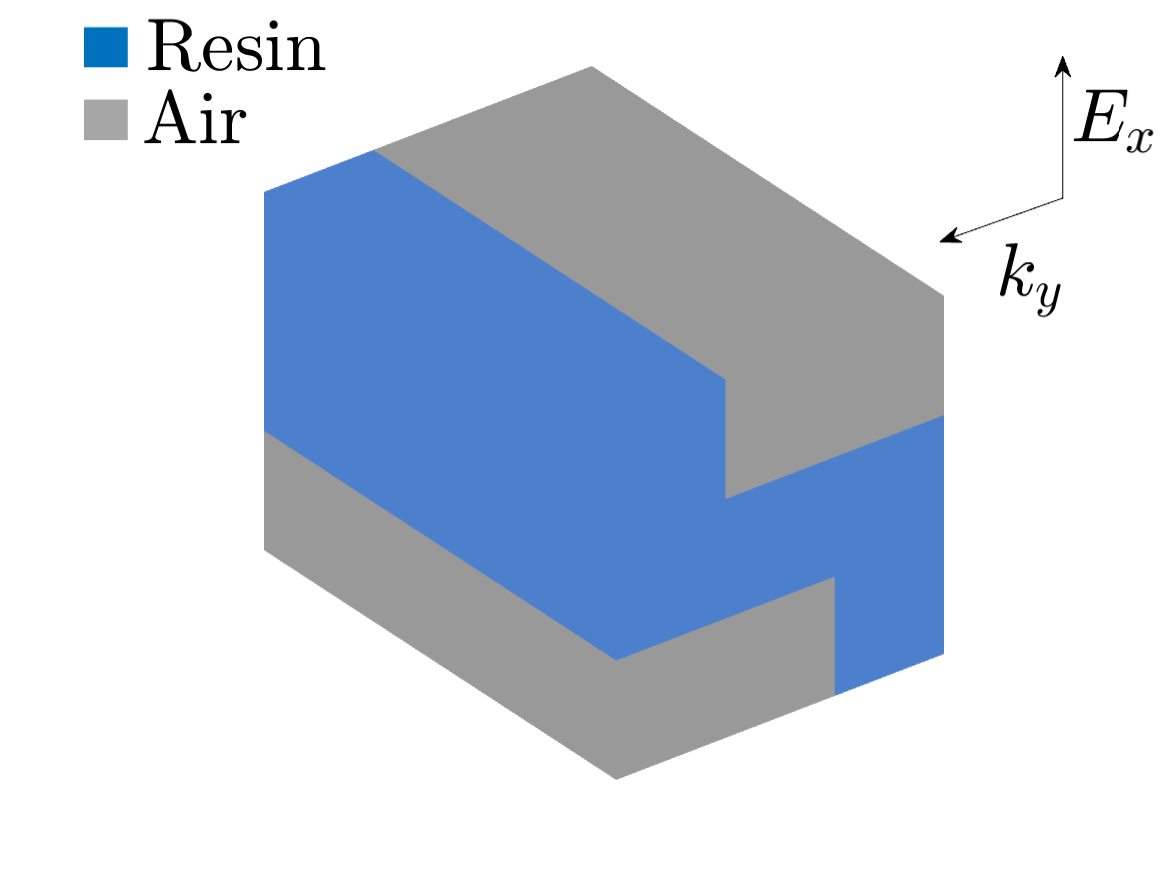}
  }
\end{overpic}
\caption{The dispersion diagram  of  the air/resin unit cell with 55.6\% resin matches that of an effective material of $\epsilon_r$=1.7.}
\label{fig:beta}
\end{figure}

\subsection{Assigning a resin percentage to each discrete material level}
We assign a resin percentage to each intermediate material level, as shown in Fig.~\ref{fig:overview}'s Table I.  Resin percentage
does not exclusively define the effective dielectric
constant. Structures of a uniform or close to uniform air/resin distribution have an effective propagation constant very close to the average of all air/resin distributions~\cite{Passia2024}. Unit cells with uniform air/resin distribution, i.e.,
finer internal structures, are more suitable to approximate the
assigned dielectric constant.
Hence, we consider close-to-uniform configurations that are also manufacturable, as is the unit cell shown as an inset of Fig.~\ref{fig:beta}, to determine the suitable resin percentage. 
The dispersion diagram of the unit cell with 55.6\% resin is depicted in Fig.~\ref{fig:beta}. The propagation constant of this unit cell matches that of a homogeneous material with $\epsilon_r$~=~1.7.
For each intermediate material level we select the resin percentage for which the propagation constant of the mode supported by the air/resin configuration matches that of the intermediate material level. 

\subsection{Identifying suitable air-resin structures locally}

The \texttt{LOCABINACONN} algorithm is implemented for each non-manufacturable material component of a dielectric constant level $\epsilon_{\rm r,d}$.   \texttt{LOCABINACONN} assigns air or resin to all grid cells of an $\epsilon_{\rm r,d}$ dielectric constant.  The components of a material level with $\epsilon_{\rm r,d}$~=~1.28 are shown in Fig.\ref{fig:overview}b, while only the largest material component is illustrated in Fig.\ref{fig:overview}c. 

The steps of the methodology are as follows:
\begin{itemize}
\item For each material level $\epsilon_{\rm r,d}$, we locate its components, i.e., material regions that are intrinsically connected.
\item  For each component $i$ of the material $\epsilon_{\rm r,d}$, we generate multiple manufacturable air/resin configurations of the same resin percentage but with  air and resin distributed differently across the component. 
\item  To generate each manufacturable air/resin configuration, we set $n_{a}$ cells to air, according to the assigned air/resin percentage. 
  \begin{itemize}
  \item A uniform random distribution is used to determine the cells that will be set to air in each component. We want air/resin to be distributed more uniformly across the component; i.e., to have configurations with finer features.  
\item Before setting a cell to air, we check that the total device remains connected upon removal of the cell. If it becomes disconnected, we return to the previous configuration.
  \end{itemize} 
\item We identify the manufacturable component that better approaches the non-manufacturable component, by comparing their S-parameters.  
\item By using the \texttt{LOCABINACONN} algorithm on all components of all non-manufacturable material levels, we obtain a manufacturable device, which is binary and connected.
\end{itemize}

To find the components of each material and check that the device remains connected upon removal of a cell, we use basic concepts of graph theory. We form a graph $G$, that consists of a set of nodes $V$, which in our case are all non-air cells ($\varepsilon_r>1$). Two nodes are connected by an edge $E$ if they are immediate neighbors (west, north, east, south). We use MATLAB's  \texttt{conncomp} function  to calculate the connected components of graph $G$,  i.e., groups of nodes that are intrinsically connected. The entire device is connected if there is a single component in the graph $G$ that contains all nodes.  We also form a graph $G_d$ that contains the cells of a material level $\epsilon_{r,d}$.  The material components of the material level $\epsilon_{r,d}$ are determined by using the same process.

 In the considered example, we generate ten manufacturable configurations for each component. The number of generated manufacturable configurations can be adjusted between components; we may generate more configurations for larger components and fewer for smaller components. As an example, we consider the largest component with $\epsilon_{\rm r,d}$~=~1.28 (Fig.\ref{fig:overview}c) and generate ten manufacturable configurations. In Fig.\ref{fig:overview}c, we present one of the ten manufacturable configurations. The non-manufacturable component is substituted by resin (black) and air (white). The remainder of the device is shown in blue. We observe that all resin parts are connected to the device, verifying that the entire device remains manufacturable.

\begin{figure}
\includegraphics[width = 0.9\columnwidth,  trim = {0cm 0cm 0cm 0cm}, clip]{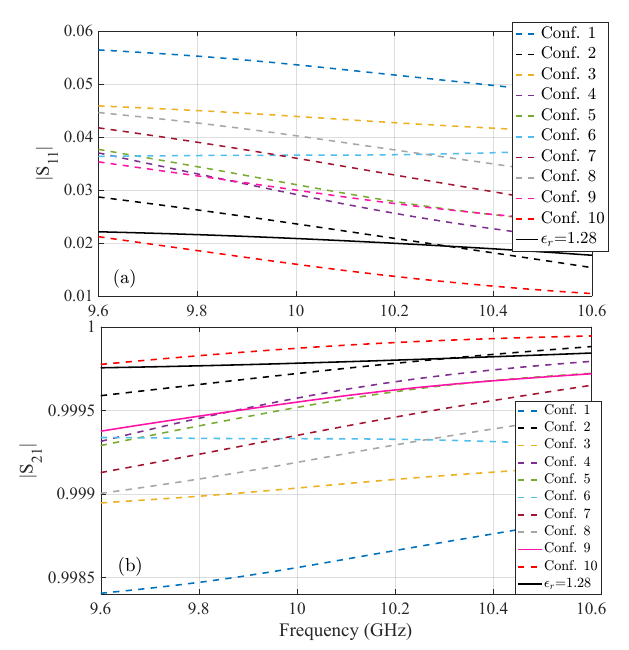}
\caption{(a) The reflection  $|S_{11}|$ and (b) transmission  $|S_{21}|$ coefficients of ten   manufacturable air/resin configurations that try to approach the non-manufacturable material component of Fig.~\ref{fig:overview} with $\epsilon_r$=1.28.}
\label{fig:all_S}
\end{figure}

We perform FEM simulations of each manufacturable component and of the corresponding non-manufacturable component.  Each component is surrounded by an appropriate computational domain with the dielectric constant of the surrounding space set to air. The FEM simulations are carried out in the RF module of COMSOL Multiphysics, with periodic boundary conditions imposed on side boundaries. The manufacturable configuration is excited by a normally incident plane wave, as in Fig.~\ref{fig:overview}.  The simulated reflection and transmission coefficients are shown in Fig.~\ref{fig:all_S}a and Fig.~\ref{fig:all_S}b, respectively, for the non-manufacturable and all ten manufacturable components.  We observe that the second manufacturable configuration better approaches the response of the non-manufacturable component.  We perform this process on all materials and material components, starting from the sparser material and moving on to the denser materials. Once we select the best matching manufacturable component, we update the device's material distribution. The binary and connected device is obtained once all non-manufacturable components are substituted by manufacturable ones.

\section{Validation and Discussion}

\begin{figure}
  \subfloat[]{\label{fig:manuf_device_local}%
  \includegraphics[width = 0.2\columnwidth,,trim={0cm 0cm 0cm 0cm},clip]{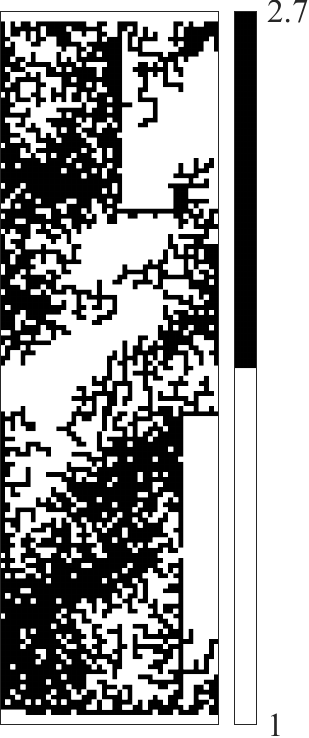}\\
     \vspace{-1.4em}
  }
  \subfloat[]{\label{fig:eff_local}%
  \includegraphics[width = 0.8\columnwidth,trim={0cm 0cm 0cm 0cm},clip]{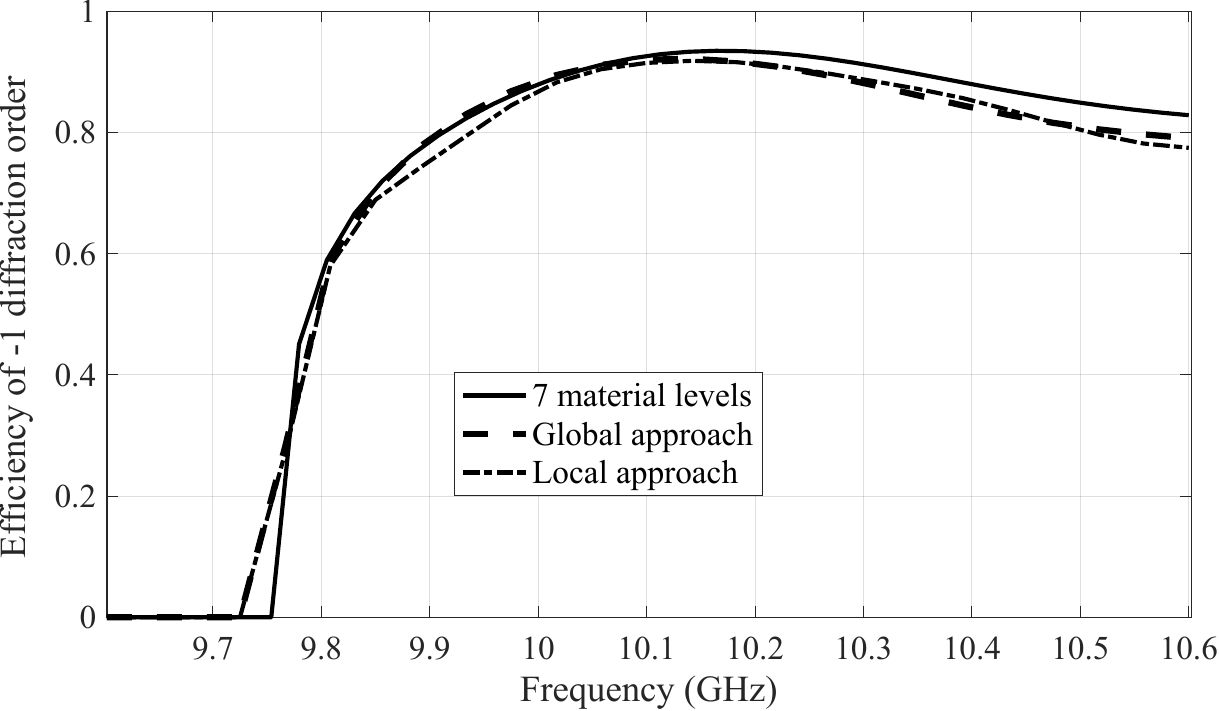}\\
    \vspace{-1.4em}
  }
\caption{(a) Manufacturable device by \texttt{LOCABINACONN}. (b) The efficiency of the $-1$-diffraction order for the manufacturable device obtained by  \texttt{LOCABINACONN} is compared to that of a device obtain by a global approach~\cite{Passia2024} and the device with seven material levels.}
\label{fig:all_eff}
\end{figure}

The 3D-printable device, composed of only air and resin, is shown in Fig.~\ref{fig:all_eff}a.   We validate our methodology by simulating the manufacturable device unit cell using the FEM solver of COMSOL Multiphysics. We compare the $-1$ diffraction order efficiency of the manufacturable device to that of the non-manufacturable device with seven materials and to the manufacturable device obtained by a global approach that requires simulations of the entire unit cell~\cite{Passia2024}. In Fig.~\ref{fig:all_eff}b, we observe that the performance of the locally designed manufacturable device is very close to that of the non-manufacturable device. The simulations provide sufficient validation for our methodology, since we
have ensured that the devices produced by \texttt{LOCABINACONN}  can be
manufactured, adhering to the Formlabs Form 3 printer  specifications. The average efficiency is 85.22\% for the locally-designed and  85.88\% for the globally-designed manufacturable devices, and 88\% for the non-manufacturable device in the \{9.9,10.6\}~GHz range. Hence, the \texttt{LOCABINACONN} methodology has a competitive performance with a global design approach, while also alleviating the need to simulate the entire manufacturable component multiple times.
Our work paves the way to synthesizing three-dimensional 3D-printable devices that retain the high performance of the inverse-design outcomes in a computationally tractable manner.

\section*{Acknowledgment}
This project has received funding from the European Union's Horizon 2020
research and innovation programme under the Marie Skłodowska-Curie grant
agreement No.101146306.
\includegraphics[width =0.5\columnwidth]{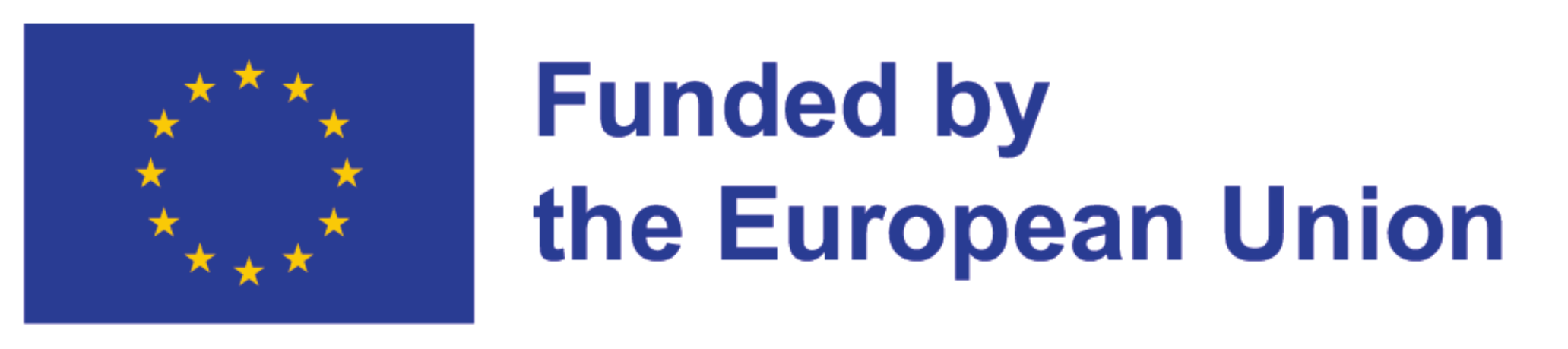}

%\section{References}

\bibliography{ref} % Entries are in the refs.bib file

\bibliographystyle{IEEEtran}

\end{document}